\documentclass{iau} 
\usepackage{graphicx}

\title[Multiple Populations in Cluster Formation Simuations] %% give here short title %%
{The Emergence of Multiple Populations in Radiation Hydrodynamics Simulations of Cluster Formation}

\author[Alison Sills]   %% give here short author list %%
{Alison Sills}

\affiliation {Department of Physics \& Astronomy, McMaster University, 1280 Main Street West, Hamilton, ON, L8S 4M1, CANADA \\ email: {\tt asills@mcmaster.ca} }
\pubyear{2019}
\volume{351}  
\jname{Star Clusters: From the Milky Way to the Early Universe}
\editors{A. Bragaglia, M.B. Davies, A. Sills \& E. Vesperini, eds.}
\begin{document}

\maketitle
%. CONTINUE EDITING FROM HERE

\begin{abstract}

We present a new approach to understanding star-to-star helium abundance variations within globular clusters. We begin with detailed radiation hydrodynamics simulations of cluster formation within giant molecular clouds, and investigate the conditions under which multiple populations could be created. Chemical enrichment occurs dynamically as the cluster is assembled. We test two extreme mechanisms for injection of enriched gas within the clusters, and find that realistic multiple populations can be formed in both mechanisms. The stochastic cluster formation histories are dictated by the inherent randomness of the timing and location of the formation of small clusters, which rapidly merge to build up the larger cluster, in combination with continual accretion of gas from the cloud. These cluster formation histories naturally produce a diversity of abundance patterns across the massive cluster population. We conclude that multiple populations are a natural outcome of the typical mode of star cluster formation. 

\end{abstract}

\firstsection % if your document starts with a section,
\section{Introduction}

Another contribution to this proceedings (Harris, this volume) describes a suite of radiation hydrodynamic simulations of the collapse of giant molecular clouds and the subsequent formation of massive star clusters. We used these simulations to investigate the imprinting of multiple population signatures on the stars in these forming clusters. Our motivation was to take a step back from the many nucleosynthetic-based scenarios, and start with something which is irrefutable: star clusters form. If we can understand that process, can we then extend our understanding to include the multiple populations question? A more detailed description of the work presented here can be found in \cite{Howard2018} and \cite{Howard2019}.

``Multiple populations" is the phrase used to capture the appearance of star-to-star variation in light elements (C, N, O, Na, Al, Mg) in massive star clusters, without an accompanying variation in iron or heavy elements. Only a few clusters show an iron spread and/or evidence of variation in r-process and s-process elements, but essentially all clusters older than about 2 Gyr and more massive than about $10^4$ M$_{\odot}$ show light element variation. For a detailed description of our understanding of the complexity of this problem, see contributions in this volume and the recent review of \cite{BL2018}. While the problem is complex, there is agreement on the following points:

\vspace{2ex}

\begin{itemize}
\item Multiple populations are primarily a ``cluster" phenomenon, in that very few field stars show these particular abundance patterns.
\item The particular abundance patterns indicate material processed through hot hydrogen burning, and therefore helium enrichment.
\item There is no observable age difference between stars of different chemical abundances. 
\item While the existence of multiple populations is ubiquitous, the amount and pattern of enrichment varies considerably from cluster to cluster. 
\end{itemize}

\vspace{2ex}

Our models of star cluster formation show that massive clusters are built up through two simultaneous processes. Small clusters form in the densest regions of the giant molecular cloud, and then grow through continued accretion of gas as well as merging with other small clusters. The most massive clusters in our simulation have approximately equal amounts of their final mass from merged clusters, and from direct gas infall. Each massive cluster also has a unique formation history, with different masses of clusters merging with the main cluster at different times, and different amounts of gas accreted by both the main cluster and the smaller clusters at different times. Giant molecular clouds are inherently stochastic objects, and that stochasticity becomes imprinted on the clusters that form. Our simulations also show that massive star clusters are built up quickly -- within 5 Myr of the start of the simulation, or 2-3 Myr after star formation begins. Therefore, our simulations naturally address the last two points in the previous list. We took our models one step further by combining the first two points to address the following question: If clusters produced significant amounts of helium-enriched material, would we see multiple populations in our simulations? 

\section{Two Extreme Enrichment Mechanisms}

\begin{figure}[h]
\vspace*{0cm}
\begin{center}
\includegraphics[width=\textwidth]{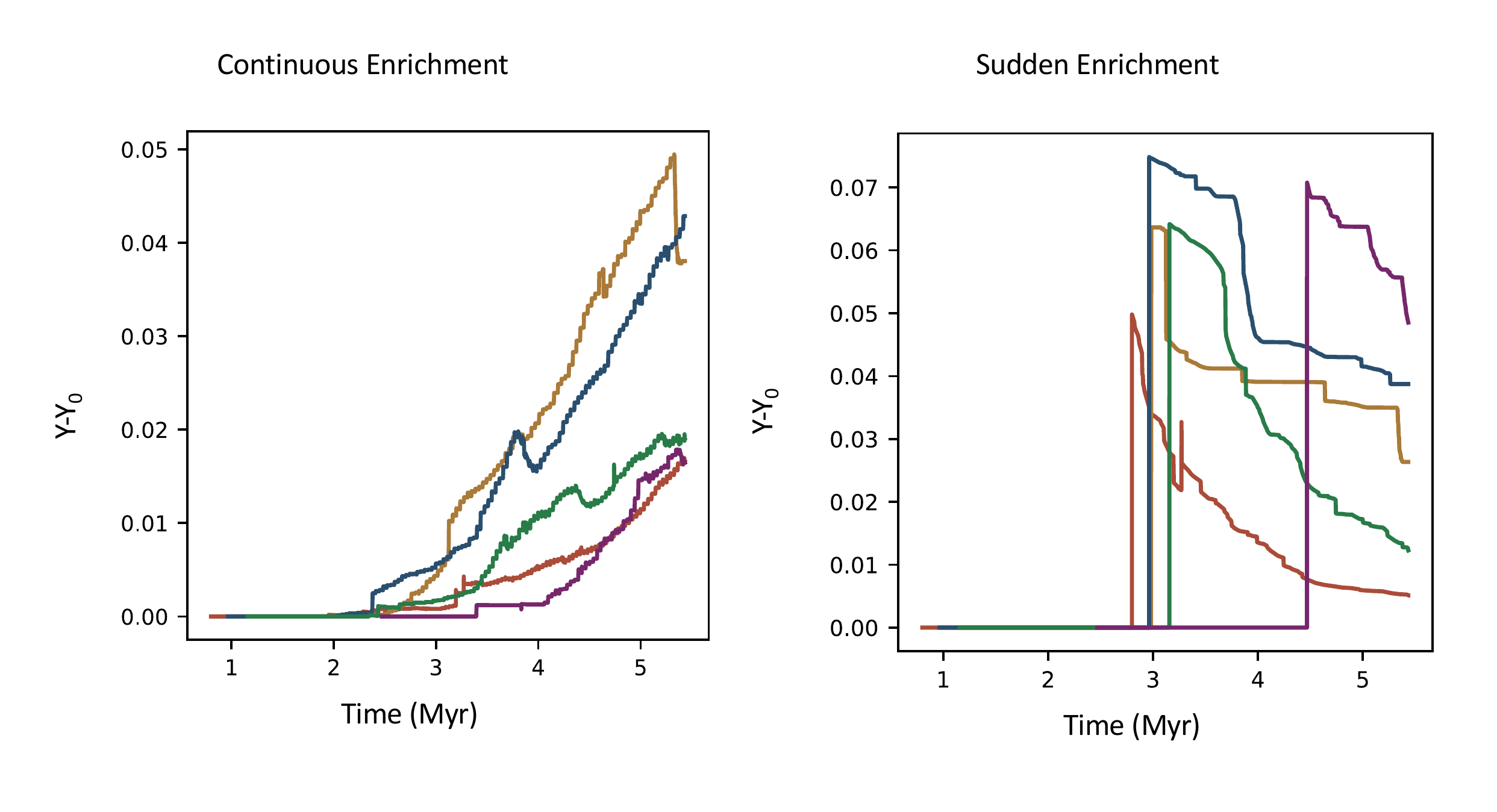}
\vspace*{0cm}
 \caption{Average helium abundance as a function of time for the 5 most massive clusters in our simulations.  In the `continuous enrichment' models, the helium abundance in the cluster gas is increased in proportion to the recent star formation rate and so in general keeps increasing with time. Reductions in average helium abundance occur when a large but relatively unenriched cluster merges with the main cluster, or when the main cluster accretes significant amounts of gas from the giant molecular cloud. On the other hand, the `sudden enrichment' models show a large increase in helium abundance 2 Myr after star formation begins in the cluster, and then the helium abundance is reduced as the clusters continue to accrete gas from the giant molecular cloud.}
   \label{fig1}
\end{center}
\end{figure}

We chose to investigate two extreme possibilities for the rate at which helium-enriched material is created inside the forming clusters. Under the assumption of `continuous enrichment', the amount of helium-rich material is tied to the star formation rate within the cluster. If instead we assume `sudden enrichment', then all the helium-rich material is introduced into the cluster at a particular time (chosen in this simulation to be 2 Myr after star formation begins in each cluster). We assume that as stars are formed in the clusters, they retain the abundance of the gas at their moment of formation. We further assume that the enrichment only happens inside clusters, and that the composition of gas in the giant molecular cloud is unchanged. Therefore, a small cluster can be enriched by either mechanism, but as it accretes more gas from the surrounding cloud, the helium abundance of the gas in the cluster can be reduced. In addition, stars in each small cluster are forming throughout the duration of simulation, and therefore are being enriched at different times. As clusters merge, their stars with their different compositions are combined into a larger cluster. This implies that there is not a direct relationship in time between the amount of enrichment and the time of formation of stars. It is entirely feasible in this model for stars with lower helium to be formed after stars with high helium. This can be seen in Figure \ref{fig1}, which shows the average helium abundance in 5 massive clusters as a function of time. The structure in the curves shown in this figure indicate when smaller clusters merged with the main cluster and when significant amounts of gas was accreted from the giant molecular cloud.

In our simulations, all massive clusters showed some level of helium enrichment by 5 Myr after the start of the collapse of the giant molecular cloud. The maximum level of helium enrichment and the fraction of stars that would be observationally identified as belonging to the enriched population encompass the observations of Milky Way globular clusters. We produce a wide variety of histograms of enrichment levels from cluster to cluster, in agreement with the diversity of populations in the observations. 

We conclude that the signatures of multiple populations can arise as a normal byproduct of cluster formation in the context of a rigorous, quantitative radiation hydrodynamics model of molecular cloud collapse. This framework does not suffer from the mass budget problem since the molecular cloud provides a reservoir of star-forming gas, and so we do not need to invoke non-standard initial mass functions for either population, nor substantial stellar mass loss from the cluster after formation. We naturally explain the variation from cluster to cluster, the lack of an age difference between the two populations, and we expect that more massive clusters should show a larger spread in helium abundance as they can more easily retain their enriched material. In addition, in our model the star-to-star abundance spreads are generated by the cluster formation process, and stars of different abundances are formed at the same time. As such, the language of `first' and `second' generations often used in the literature is misleading and incorrect. 

\section{Future Work}
 
The astute reader will notice that we have not identified the stellar object(s) responsible for the hot hydrogen burning that created the helium-enriched material in either the sudden or continuous enrichment models. For this first step, we wanted to explore the possibilities with these simplified formulations. But both mechanisms were prompted by possible nucleosynthetic sources. The continuous enrichment model used the total mass of massive stars to determine the amount of helium that was added to the cluster. Massive stars in interacting binaries have been shown to produce abundance patterns that are similar to those seen in the enriched globular cluster stars (\cite{deMink2009}). The sudden enrichment model was prompted by the suggestion that supermassive stars could be produced in young dense clusters (\cite{Gieles2018}). Both massive binaries and supermassive stars have the benefit that both require a star cluster to maximize the amount of helium-enriched material that can be recycled before helium-burning occurs. Supermassive stars are thought to form through the collision of regular massive stars (\cite{SillsGlebbeek2010}), which requires the dense environment of a cluster to provide a high enough collision rate. Massive stars are more likely to be in binaries than solar-type stars, and more likely to be in close binaries with other massive stars (\cite{MoeDiStefano2017}). In clusters, dynamical interactions mean that hard binaries will get harder (Heggie's Law, \cite{Heggie75}) and so massive binaries in clusters are more likely to be forced into separations where the kinds of binary evolution described in \cite[de Mink et al. (2009)]{deMink2009} are more likely to occur. A more in-depth investigation of the yields from a population of massive binaries is needed. 

In addition, we are following up this work with simulations that track the stellar dynamics of a dense cluster environment in the presence of significant amounts of gas. It may be that both massive binaries {\it and} supermassive stars contribute to the enrichment of the gas in the cluster. We also need to include a proper treatment of the hydrodynamics of the gas coming into the cluster from the molecular cloud, in combination with the hydrodynamics of the stellar winds. These simulations will need to be at a higher resolution than the cluster formation simulations described above in order to resolve the interesting physical processes. We envision using the large-scale simulations as external boundary conditions and initial conditions for the detailed analysis of individual clusters. The combination of the multi-scale, multi-physics approach will provide a powerful tool to understand star cluster formation and early evolution, including but not limited to the multiple population problem.

\end{document}